\documentclass[12pt]{article}

\usepackage{tikz}
\usepackage{epsfig,latexsym,amsfonts,amsmath,amsthm,amssymb,amsbsy,multirow,slashed,wasysym,textcomp,subfigure,wrapfig,datetime,comment,mathtools,cancel,cite,twistor,mathrsfs}
\usepackage[hidelinks]{hyperref}
\def\ee{\end{equation}}
\def\be{\begin{equation}}
\def\bea{\begin{eqnarray}}
\def\eea{\end{eqnarray}}
\newcommand{\beq}{\begin{eqnarray}}
\newcommand{\eqq}{\end{eqnarray}}
 \newcommand{\badat}{\begin{alignedat}}
 \newcommand{\eadat}{\end{alignedat}}

\newcommand{\eal}[1]{\be \begin{aligned} #1 \end{aligned}\end{equation}} 
\newcommand{\eqn}[1]{\be #1 \end{equation}} 
\newcommand{\eqa}[1]{\bea  #1\end{eqnarray}}

\renewcommand{\d}{\mathrm{d}}

\long\def\new#1\endnew{{\bf #1}}		
\long\def\del#1\enddel{}
\def\eps{\epsilon }
\def\del{\partial}

\usepackage{color}

\newcommand{\pink}[1]{\textcolor{\pink}{#1}}

\definecolor{dblue}{rgb}{0.2,0.50,0.80}

\def\et2{ET$^2$}
\def\m2{M$^2$}

\def\O{\mathcal{O}}

\def\D{{\Delta}}

\def\sllr{$\mathrm{SL}(2,\mathbb{R})\otimes \overline{\mathrm{SL}}(2,\mathbb{R})$}

\def\bigma{\bar\sigma}
\def\e{{\epsilon}}

\def\bz{{\bar z}}

\def\s{ {\sigma} }
\def\co{{\cal O}}
\newcommand{\bs}{\bar{\sigma}}

\numberwithin{equation}{section} % equation numbers follow sections

\begin{document}
\begin{titlepage}
  \thispagestyle{empty}
  \begin{flushright}
%eDRAFT~~~  \today
    \end{flushright}
  \bigskip
  \begin{center}
	 \vskip2cm
  \baselineskip=13pt {\LARGE \scshape{Conformal Field Theory with Periodic Time  %\\
  %\vspace{0.5em}
}}

	 \vskip2cm
   \centerline{Walker Melton and Andrew Strominger}
 \vskip.5cm
 \noindent{\em Center for the Fundamental Laws of Nature,}
  \vskip.1cm
\noindent{\em  Harvard University,}
{\em Cambridge, MA, USA}
\bigskip
  \vskip1cm
  \end{center}
  \begin{abstract}
  
  It is shown that time-ordered correlation functions  of a unitary CFT$_2$ in 2D Minkowski space admit  a single-valued, conformally-invariant  extension to the   Lorentzian signature torus provided that the $S^1\times S^1$ spatial and temporal radii are equal.  The result  extends to Lorentzian CFT$_D$ on equal-radii $S^{D-1}\times S^1$ under the assumption that branch cuts  occur only when a pair of operator insertions are null separated. 
  \end{abstract}
% \noindent
%	

%
\end{titlepage}
\tableofcontents

\section{Introduction}

Lorentzian conformal field theory in $D$ dimensions (CFT$_D$)  is often studied in Minkowski space M$^D$ or the Einstein cylinder $S^{D-1}\times R$.  In either case, the correlators do not admit a canonical  action of the full conformal group SO$(D,2)$. The boundary at infinity of  M$^D$ is preserved only by the Poincare-dilational subgroup, while correlators on the  Einstein cylinder EC$^D$ are  acted on by the universal cover of SO$(D,2)$  \cite{Kravchuk:2018htv}.  The Lorentzian spacetime which does admit a good SO$(D,2)$ action is the equal-radii $S^{D-1}\times S^1$ `Einstein torus', denoted ET$^D$.

 Sections 2-5 of this paper concern the case $D=2$. We show  that the time-ordered correlators in a unitary CFT$_2$ with integral spins  in M$^2$ have a single-valued analytic continuation to ET$^2$, where they admit an action of SO$(2,2)$.\footnote{ Half-integral spins may also be considered but it would  require specification of a spin structure.} The simple demonstration herein amounts, in the 2D case, to replacing the \m2\ coordinates $t^\pm$  with  \be \sigma =\arctan t^+,~~~\bigma=\arctan t^- ,\ee
allowing $(\sigma,\bigma)$ to range over \et2\ and checking the branch cut behavior.\footnote{ The general form of single-valued correlators on ET$^2$ was found in  \cite{Melton:2023hiq}. } Through three points invariance is manifest in direct inspection. At four and higher points monodromy of combinations of conformal blocks must be  considered, but the constraints required by single-valuedness of the usual time-ordered product  on EC$^2$ turn out to guarantee it for our extension to   \et2. As we see in section 2.2 the Wightman functions in M$^2$ do {\it not} admit a single-valued continuation to \et2. 

 At least through three points our construction seems almost trivial  from these  perspectives, yet it is  surprising from others. First, it is generically difficult to define QFT on any spacetime with closed timelike or null curves. Second, for a generic unitary CFT$_2$, states in highest weight representations do not have integral energies and so are not periodic under time evolution.\footnote{In \cite{Melton:2025ecj} it was shown in a holographic context that the states can be thought of as lying in  the principal series whose states are  manifestly time-periodic.  }  Nevertheless we will see herein %by direct computation 
 that these observations do not obstruct the existence of well-defined correlators on the Einstein torus transforming in representations of the conformal group.

We extend this result to $D$ dimensions in section 6. Here the embedding formalism in which spacetime is a projective  section of the lightcone in M$^{D,2}$ is useful.  ET$^D$ is a global section, while M$^D$ or EC$^D$  cover only certain patches. We make  the strong  assumption - known to hold only for $D=2$ \cite{Maldacena:2015iua} - that  branch points of the time ordered n-point function arise only when a pair of operators are null separated. Under this assumption, it is shown that time-ordered correlators of a unitary CFT$_D$ for any $D$  admit an  extension from M$^D$ to ET$^D$. We first show that, when an operator is dragged around any cycle,  the extension of the time-ordered cross-ratios to ET$^D$ may  cross the real axis  only at one value. It then follows that  there is no monodromy in their  phases and the correlator is single-valued as a function of each of the operator insertions.\footnote{This analysis provides an alternate proof of the 2D case.} %We further describe in detail how the phases acquired at lightcone crossings always conspire to cancel.     

Sections 6 closes with a discussion of possible implications of our results.

\section{Two points}
In this section we construct the two point function on \et2. 
Consider a generic unitary CFT$_2$ on the 2D Minkowski plane with line element
\be ds^2=-4dzd\bz, \ee
where 
\be z\sim {\tau+x\over 2},~~~~~\bz\sim {\tau-x\over 2}\ee are both  future-increasing null coordinates (and the bar is not complex conjugation).
The CFT$_2$ is characterized by primary operators $\cO_k(z_k,\bz_k)$ with weights and spins $(\D_k,J_k)$ obeying $J_k \in {\mathbb Z}$ along with  a collection of correlation functions. 
\subsection{Time-ordered correlator}
We normalize so that the the {\it{time-ordered}}  two point function on the initial Minkowski region, denoted   M$^2_I$,  is 
\bea \label{cm2}&& \langle\cO^{\D,J}(z_1,\bz_1)\cO^{\D,J}(z_2,\bz_2)\rangle_{TO} =\Theta (\tau_{12})\langle \cO^{\D,J}(z_1,\bz_1)\cO^{\D,J}(z_2,\bz_2)\rangle_W +\Theta (\tau_{2 1})\langle \cO^{\D,J} (z_2,\bigma_2)\cO^{\D,J} (z_1,\bz_1\rangle_W \cr &&
~~~~~~~~~~~~~~~~~~~~~~~~~~~~~~~~~~~~={1\over (i\e-z_{12}\bz_{12})^{\D}}\left({  \bz_{12}-i\e z_{12}\over z_{12}-i\e\bz_{12}}\right)^{J},\eea
where $z_{12}=z_1-z_2$, $(\D_1,J_1)=(\D_2,J_2) $, $\langle...\rangle_W$ is the Wightman function  and we define the branches according to 
\be\label{bcp}
\lim_{\eps\to0}\,{1 \over (x+\im\eps)^{\D }}= \begin{cases}
{1 \over |x|^{\D}}\qquad &x>0\\
{e^{-\im\pi \D} \over |x|^\D}\qquad &x<0
\end{cases}
\ee
The $i\e$ prescription here is the standard one inherited from Euclidean space.\footnote{See sections of \cite{Hartman:2015lfa,HarrisSimmonsDuffinTASI,Kundu:2025jsm} for  reviews. Note that 
in our conventions $z\bz\sim {(x^0)^2-(x^1)^2 \over 4}$ and hence is negative for  spacelike separations.} We do not determine codimension 2 contact terms $e.g.$ where $z_{12}=\bz_{12}=0$ in this paper.

To extend this to the Einstein torus \et2\, define
\be \label{zsig}
z = \tan\sigma\,,\qquad\bz = \tan\bigma
\ee
where \be\label{shft}
(\sigma,\bigma) \sim \big(\sigma+(m+n)\pi,\,\bigma+(m-n)\pi\big)\,,\qquad m,n\in\Z\,.
\ee
In terms of 
\be \sigma={t+\phi \over 2}, ~~~~\bigma={t-\phi \over 2},\ee 
this is equivalent to $2\pi$ periodicity of $t$ and $\phi$.
A convenient fundamental domain for these coordinates can be taken to be
\be \label{fdm} 
0\leq\sigma<2\pi\,,\qquad0\leq\bigma<\pi.
\ee
\begin{figure}
\begin{center}

\tikzset{every picture/.style={line width=0.75pt}} %set default line width to 0.75pt        

\begin{tikzpicture}[x=0.75pt,y=0.75pt,yscale=-1,xscale=1]
%uncomment if require: \path (0,235); %set diagram left start at 0, and has height of 235

%Shape: Square [id:dp294442792323391] 
\draw   (173,77.64) -- (240.4,145.04) -- (173,212.44) -- (105.6,145.04) -- cycle ;
%Shape: Square [id:dp03571715654511698] 
\draw   (240.4,10.25) -- (307.79,77.64) -- (240.4,145.04) -- (173,77.64) -- cycle ;
%Straight Lines [id:da32380178252177394] 
\draw    (99,153) -- (162,217.44) ;
\draw [shift={(125.96,180.57)}, rotate = 45.65] [fill={rgb, 255:red, 0; green, 0; blue, 0 }  ][line width=0.08]  [draw opacity=0] (8.93,-4.29) -- (0,0) -- (8.93,4.29) -- cycle    ;
%Straight Lines [id:da689853488678803] 
\draw    (315,84.56) -- (184,218.44) ;
\draw [shift={(254.05,146.85)}, rotate = 134.38] [fill={rgb, 255:red, 0; green, 0; blue, 0 }  ][line width=0.08]  [draw opacity=0] (8.93,-4.29) -- (0,0) -- (8.93,4.29) -- cycle    ;
%Straight Lines [id:da6527794559688003] 
\draw    (120,122) -- (136,122.44) ;
%Straight Lines [id:da29302208109585737] 
\draw    (261,117) -- (277,117.44) ;
%Straight Lines [id:da10624887305517472] 
\draw    (194,50) -- (210,50.44) ;
%Straight Lines [id:da9028947447004533] 
\draw    (198,45) -- (214,45.44) ;
%Straight Lines [id:da5176101630465916] 
\draw    (194,182) -- (210,182.44) ;
%Straight Lines [id:da7469608712858289] 
\draw    (198,177) -- (214,177.44) ;
%Curve Lines [id:da733876302675966] 
\draw    (265,44) .. controls (277,29.44) and (273,52.44) .. (281,38.44) ;
%Curve Lines [id:da35225481761987143] 
\draw    (133,180) .. controls (145,165.44) and (141,188.44) .. (149,174.44) ;

% Text Node
\draw (255,148) node [anchor=north west][inner sep=0.75pt]    {$\sigma $};
% Text Node
\draw (111,189) node [anchor=north west][inner sep=0.75pt]    {$\overline{\sigma }$};

% Text Node
\draw (165,138) node [anchor=north west][inner sep=0.75pt]    {M$^2_I$};

% Text Node
\draw (225,75) node [anchor=north west][inner sep=0.75pt]    {M$^2_{II}$};

\end{tikzpicture}
\end{center}
\caption{\label{fig:celtor}A  fundamental domain in $(\sigma,\bigma)$ for the Einstein torus \et2 . The two diamonds  M$^2_I$ and M$^2_{II}$ are each Weyl-equivalent  to 2D Minkowski space \m2 . }
\end{figure}
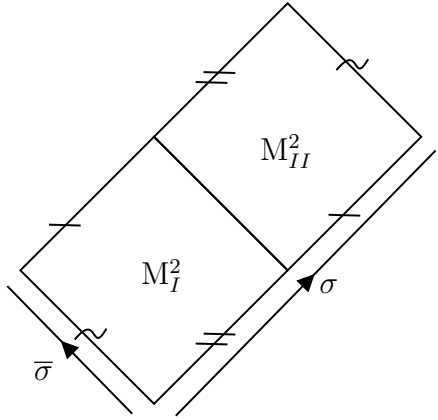
The $(z,\bz)$ coordinates cover only half of this region where $-\pi/2 < \s,\bs < \pi/2$. 
\et2\ is the  union of two antipodally placed Minkowski  diamonds  M$^2_I$ and M$^2_{II}$. On each diamond, one can introduce coordinates in which, after 
 Weyl rescalings, the corresponding 2D metric is $-4 \d z\,\d\bz$. 
 The 2D conformal group $\SL(2,\R)\times\overline{\SL}(2,\R)$ acts by real and independent M\"obius transformations on $z$ :
\begin{equation}
    z \mapsto \frac{az+b}{cz+d}\,,\qquad \s \mapsto \arctan\left(\frac{a\tan\s+b}{c\tan\s+d}\right)
\end{equation}
where $ad-bc = 1$, with a similar formulae for $\bz,~\bs$. 

We wish to continue the \m2\  correlator \eqref{cm2} to \et2. One finds using the identity 
\be
z_{12} = \frac{\sin\sigma_{12}}{\cos\sigma_1 \cos\sigma_2}\,,\qquad\bz_{12} = \frac{\sin\bigma_{12}}{\cos\bigma_1 \cos\bigma_2}\,.
\ee
and the conformal transformation properties \be \cO^{\D,J}(\sigma,\bigma) = |\cos\sigma|^{-\D-J}|\cos\bigma|^{-\D+J}\cO^{\D,J}(z,\bz)\, \ee  that \eqref{cm2} may be continued to 
\be\label{cet2} \langle \cO^{\D,J}(\sigma_1,\bigma_1)\cO^{\D,J}(\sigma_2,\bigma_2)\rangle_{ET_2} = {1\over (i\e-\sin\sigma_{12}\sin\bigma_{12})^{\D}}\big({ \sin\bigma_{12}-i\e\sin\sigma_{12} \over \sin\sigma_{12} -i\e \sin\bigma_{12}}\big)^{J}.\ee
This is invariant under the identifications \eqref{shft} acting on either $\sigma_1$ or $\sigma_2$. One must additionally  check that orbits of an operator insertion  around the nontrival cycles of the torus do not impart phases to the correlators. %
To drag $\cO_1$ around the timelike circle (with $\cO_2$ held fixed) one shifts both $\sigma_{12}$ and $\bigma_{12}$ both by ${t\over 2}$ and takes $t$ from  $0$ to $2\pi$. For generic initial positions, this trajectory crosses two branch cuts, one  with $\sigma_{12}=0$ and one with $\bigma_{12}=0$, and necessarily with different initial signs of 
 $\sin \sigma_{12}\sin \bigma_{12}$ so that the phases the two-point function develops when $\O_1$ crosses $\O_2$'s lightcone exactly cancel. Hence the correlator is single-valued.
  
 This is closely related to the well-known fact that \m2\ correlators admit an extension to the Einstein cylinder which has no monodromy around the spatial circle.\footnote{In turn related to the single-valuedness of Euclidean correlators under $2\pi$ rotations.}  Indeed the exchange $t\leftrightarrow \phi$ in \eqref{cet2} simply  complex conjugates the correlators and multiplies them by a constant phase. Hence $t$ and $\phi$ cycle monodromies are complex conjugates. 
 
 We conclude that \eqref{cet2} defines a consistent extension of time-ordered two-point CFT correlators from \m2\ to \et2\ .

In the initial $M^I_2$ correlator \eqref{cm2}, all singularities  locally take the conventional form $(z\bz-i\e)^{-\D}$.  However, in the extension to \et2\ there are additional singularites for  antipodally- located operators in different Minkowski diamonds with $\sigma_1=\sigma_2$ but $\bigma_1=\bigma_2+\pi$.\footnote{In this configuration each operator lies on a  caustic of the other's light cone.} Locally these antipodal singularities behave with the opposite $i\e$ prescription as $(z\bz+i\e)^{-\D}$. %This can be interpreted as time running backwards in $M_2^{II}$. 
%With generic locations for operator insertions there will be both types of singularities in all \m2\ diamonds.\footnote{We insert here the following  heuristic
%remark.  Let us formally consider  the operation $S^I$ ($S^{II}$) which is a map on states from $\scri^-$ to $\scri^+$ on $M_2^I$ ($M_2^{II})$. Since $\scri^+(M_2^{II})$ is the same as $\scri^-(M^I)$, and the final data on $M_2^I$ is the initial data on  
%$M^{II}_2$, we have $S^{II}\sim (S^I)^\dagger$ and $(S^I)^\dagger S^I=1$. If we had a good notion of states on slices of \m2\ in this context, the existence of the theory on \et2\ might  be related to unitarity of the theory on \m2 . }

\subsection{The cylinder}
In this section we compare and contrast the discussion here with the usual extension of CFT$_2$ to the cylinder, for which the only identification is the spatial one $\phi\sim \phi+2\pi$ or 
\be (\sigma, \bigma)\sim  (\sigma+\pi, \bigma-\pi)\ee The usual scalar Wightman function on the cylinder is obtained by shifting $t_1$, or equivalently both $\sigma$ and $\bigma$,  by $-i\e$\cite{Hartman:2015lfa,HarrisSimmonsDuffinTASI,Kundu:2025jsm}. One has for appropriate normalization (taking $J=0$)
\be\label{wet2} \langle \cO^\D(\sigma_1,\bigma_1)\cO^\D(\sigma_2,\bigma_2)\rangle_W = {1\over \big(\sin(i\e-\sigma_{12})\sin(i\e-\bigma_{12})\big)^\D}.\ee
The light cones along which this diverges tesselate the cylnder into an infinite sequence of \m2\ diamonds.
 Consider the null line $L$ of fixed positive  $\bigma_{12}={\pi \over 2}$ which snakes up around the cylinder. The Wightman function  has singularities along this line at \be \sigma_{12}= \pi n+i\e,~~~~n\in  Z \ee   whose branch cuts we take  to go up to $+i\infty$.  Moving forward in time  along $L$, each time such a singularity is passed 
 an extra phase $e^{-i\pi\D}$ is acquired according to \eqref{bcp}.  If we take the phase to be zero in the diamond `M$^2_0$' around $\sigma_{12}={\pi \over 2},~~\bigma_{12}=-{\pi \over 2}$ it will be $e^{-in\pi\D}$ in the \m2\ diamond  around $\sigma_{12}=n+{\pi\over 2},~~\bigma_{12}=-{\pi \over 2}$.\footnote{ One may  check that this is consistent for each diamond with the  phases acquired along a trajectory of varying $\bigma_{12}$ rather than $\sigma_{12} $. } Clearly the Wightman function is not periodic in time. 
 
 Now lets consider the time-ordered product\footnote{Note that the $\Theta$ functions affect the singularities only at $t_2=t_2$, but at  $t_1=t_2+n\pi$ for $n\neq 0$ they are constant.}
\bea\label{weat2}\langle \cO^\D(\sigma_1,\bigma_1)\cO^\D(\sigma_2,\bigma_2)\rangle_{TO}=\Theta (t_{12})\langle \cO^\D(\sigma_1,\bigma_1)\cO^\D(\sigma_2,\bigma_2)\rangle_W +\Theta (t_{2 1})\langle \cO^\D(\sigma_2,\bigma_2)\cO^\D(\sigma_1,\bigma_1\rangle_W .\eea
In the Minkowski diamond M$^2_0$, this reduces to
\be\label{west2}\langle \cO^\D(\sigma_1,\bigma_1)\cO^\D(\sigma_2,\bigma_2)\rangle_{TO}= {1\over \big(i\e-\sin\sigma_{12}\sin\bigma_{12}\big)^\D}, ~~~~0<\sigma_{1},\sigma_{2},\bigma_{1},\bigma_{2}<\pi. \ee
In this case, the phase decreases indefinitely in succesive  diamonds both to the past and the future and again is not periodic. 

Another \sllr\  invariant 2-point function, distinct  from both of these,  can be defined by 
\be\label{west2}\langle \cO^\D(\sigma_1,\bigma_1)\cO^\D(\sigma_2,\bigma_2)\rangle_{\rm periodic}= {1\over \big(i\e-\sin\sigma_{12}\sin\bigma_{12}\big)^\D}, ~~~~\forall(\sigma_{12},\bigma_{12}). \ee
In terms of the Wightman function
\bea\label{weat2}\langle \cO^\D(\sigma_1,\bigma_1)\cO^\D(\sigma_2,\bigma_2)\rangle_{\rm periodic}= \Theta (\sin t_{12})\langle \cO^\D(\sigma_1,\bigma_1)\cO^\D(\sigma_2,\bigma_2)\rangle_W +\Theta (\sin t_{21})\langle \cO^\D(\sigma_2,\bigma_2)\cO^\D(\sigma_1,\bigma_1\rangle_W .\eea
This agrees with the time-ordered correlator when restricted to M$_0^2$. 
Singularities near the null line  $L$ are encountered at
\be \sigma_{12}= n\pi+(-)^ni\e,~~~~n\in  Z. \ee
For even (odd)   $n$, the branch cuts are taken up (down) in the imaginary plane. The phases 
$e^{(-)^ni\pi \D}$ alternate and the two point function is periodic along L. We may therefore take a quotient and define a conformally covariant correlator on \et2\ which is given by \eqref{cet2}.

\section{Three points}
For spinless weight $\D_k$  operators $\cO_k$ the \m2\ the time-ordered three point function is 
\bea \label{cm3}&&\langle \cO_1(z_1,\bz_1)\cO_2(z_2,\bz_2)\cO_3(z_3,\bz_3)\rangle_{M_2} = \cr &&~~~~~~~~~~~~~{C_{123}\over (i\e-z_{12}\bz_{12})^{\D_1+\D_2-\D_3 \over 2}(i\e-z_{23}\bz_{23})^{\D_2+\D_3-\D_1 \over 2} (i\e-z_{31}\bz_{31})^{\D_3+\D_1-\D_2 \over 2}}.~~~~~~~~~~~~~~~~~~~~~~~~~~~~\eea
This extends to the \et2\ three point function 
\bea \label{cm5}&&\langle \cO_1(\sigma_1, \bigma_1)\cO_2(\sigma_2,\bigma_2)\cO_3(\sigma_3 \bigma_3)\rangle_{ET_2} 
= \cr&&~~~{C_{123}\over (i\e-\sin \sigma_{12}\sin \bigma_{12})^{\D_1+\D_2-\D_3 \over 2} (i\e-\sin \sigma_{23}\sin \bigma_{23})^{\D_2+\D_3-\D_1 \over 2} (i\e-\sin \sigma_{31}\sin \bigma_{31})^{\D_3+\D_1-\D_2 \over 2}}.~~~~~~~~~~~~~~~~~~~~~~~~~~~~\eea
This is  invariant under \eqref{shft} and has no monodromy around cycles of \et2\, as is the case for  the spinning generalizations. The argument follows that given for the two point function.

\section{Four points}
The  four  point function has a kinematic factor generalizing \eqref{cm3} multiplied by a  function $G(r,\bar r)$ of the \sllr - invariant cross ratios $r$ and $\bar r$. The kinematic factor is a product of two point functions and the extension from \m2\ to \et2\ gives a single-valued expression just as it did for 2 and 3 points. More consideration is needed for possible monodromies of $G$. In Euclidean space it has an expansion in conformal blocks 
\be G(r,\bar r)=\sum_{ij}G^i(r)P_{ij}G^j(\bar r),\ee
with $r^*=\bar r$. In general $r$ has monodromy around $r=0,1, \infty$ and the conformal blocks have branch cuts at these points.  The matrix $P_{ij}$ for a local CFT$_2$  is  highly constrained  by the absence of any monodromy in $G$. The Euclidean correlation functions can be Wick rotated to time-ordered correlators \cite{Luscher:1974ez}  on \m2\ using 
\be r={(z_{12}-i\e \bz_{12})(z_{34}-i\e \bz_{34})\over (z_{13}-i\e \bz_{13})(z_{24}-i\e \bz_{13})},  ~~~\bar r={(\bz_{12}-i\e z_{12})(\bz_{34}-i\e z_{34})\over (\bz_{13}-i\e z_{13})(\bz_{24}-i\e z_{13})}.\ee
as well as the Einstein cylinder. In these cases $r$ and $\bar r$ become  independent real variables. The resulting Feynman $i\e$ prescription for crossing light-cone singularities ensures that the $G$ is single-valued on \m2 when points are dragged around one another. Wick rotation coupled with a conformal transformation  further leads to single-valued correlators on the Einstein cylinder EC$_2$, including when an operator is dragged around a nontrivial spatial cycle \cite{Luscher:1974ez}. Wightman correlators are also single-valued \cite{Luscher:1974ez}.  Neither of these sets of correlators are time-periodic. 

Here we wish to find an extension from \m2\ to 
 the Einstein torus.  $(r,\bar r)$ are defined on all of \et2\ by simply inserting \eqref{zsig} into all expressions. One finds on \et2 that $r$ continues to
\be \label{rdef}
r= {(\sin \sigma_{12}-i\e\sin  \bigma_{12})\over (\sin \sigma_{13}-i\e \sin \bigma_{13})}{(\sin \sigma_{34}-i\e \sin \bigma_{34})\over (\sin \sigma_{24}-i\e \sin \bigma_{24})} ,\ee 
while $\bar r$ is given by 
\be \label{brdef} \bar r  = {(\sin \bigma_{12}-i\e\sin  \sigma_{12})\over (\sin \bigma_{13}-i\e \sin \sigma_{13})}{(\sin \bigma_{34}-i\e \sin \sigma_{34})\over (\sin \bigma_{24}-i\e \sin \sigma_{24})} .\ee
This agrees with the Feynman prescription in the \m2\ patch covered by $(z_k,\bz_k)$, but differs from the  non periodic $i\e$ prescription of the usual non-periodic extension  to EC$_2$. It obeys 
\be \bar r(\e)=r^*({1 \over \e}),\ee 

It is easy  to see that $r$ and $\bar r$ return to their  to their initial positions under $2\pi $ shifts of either $t_k$ or $\phi_k$ and so \eqref{rdef}\eqref{brdef} are themselves  single-valued functions on \et2. 
In principle $G(r,\bar r)$ might involve generic functions with  nonintegral powers of $r$ at its zeros or poles, potentially leading to nontrivial monoodromy of the $4$-point functions as one of the operators is dragged around another operator, the antipode of another operator or around  closed timelike or closed spacelike cycles on \et2.  These monodromies are highly constrained by the fact that $G$ has a single-valued extension from \m2\ to EC$^2$. 
 
 Consider 3 points at generic  locations on ET$^2$, $i.e$ not positioned at coincident or  antipodal light cone singularities of one another.  Let $S_k$ ($T_k$) denote the \m2\ diamonds  whose points are spacelike (timelike) separated from $(\s_k,\bs_k)\equiv P_k $. If all three points are  spacelike (timelike) separated from one another, then they all lie in or on the boundary of $S_1$, $S_2$ and $S_3$ 
($T_1$, $T_2$ and $T_3$). On the other hand if $P_1$ and $P_2$ are spacelike (timelike) separated from one another but timelike (spacelike) separated from $P_3$, they are all in $T_3$ ($S_3$). For generic positions, we can slightly move the diamond so that all three points are within (and not on the boundary of) a diamond.  

In conclusion, however the three $P_k$ are located within \et2, they are always contained within a common \m2\
diamond which we shall denote M$^2_{123}$.

We have defined our extension of three point correlators from \m2\ to \et2\ so that for all points within  M$^2_{123}$ they coincide with the standard 
ones on the Einstein cylinder EC$^2$. If we now add a fourth point at a $P_4$  within M$^2_{123}$, the result will be (conformal to) the standard \m2\ time-ordered four point function.   In particular there will be no monodromy as we move $P_4$ around any of the operator insertions at $P_k$. 

The difference arises when we take  the fourth point out of  M$^2_{123}$ approaching  the antipode of one of the $P_k$ where the light cone reconverges $i.e. ~~\s_k \to \s_k+\pi$.\footnote{In the standard continuation to EC$^2$, this singularity is regulated with the  Wightman $i\e$ prescription. In contrast, our continuation to \et2\ regulates it with the complex conjugate of the Feynman prescription for time ordered correlators, which is consistent with time periodicity.}
It follows from expressions  \eqref{rdef} and \eqref{brdef} for $r$ and $\bar r$ that the monodromy of $G$ as $P_4$ is taken around an antipode to $P_k$  is the inverse of the monodromy around $P_k$  itself. 
Consider $e.g.$ taking $P_4$ in a small contour around $P_3$. Then $r$ is a constant times $\s_{34}-i\e\bs_{34}$, while $\bar r$ is a constant times 
$\bs_{34}-i\e\s_{34}$. On the other hand near the antipodal point, $\sigma_4$ is near $\sigma_3 + \pi$,  $r$ is the same constant times $(\s_{34}-\pi)+i\e\bs_{34}$ while $\bar r$ is the same constant times $\bs_{34}+i\e(\s_{34}-\pi)$. This  leads to the inverse monodromies.  The constraint on $G$ which makes time-ordered correlators single-valued  as $P_4$ is taken around $P_k$ hence also guarantees that, with our $i\e$ prescription in Equations \eqref{rdef} and \eqref{brdef},  it is single-valued  as $P_4$  is taken around the antipode of $P_k$ in \et2 .

It remains to consider what happens as $P_4$ winds  around a non-trivial spatial or temporal cycle. There are special spatial cycles which wind around EC$^2$ while remaining everywhere within M$^2_{123}$, crossing the left spatial infinity and reentering the right.  Single-valuedness of $G$ around such cycles follows immediately from  single-valuedness on EC$^2$ for either $\eps>0$ or $\eps < 0$. More general winding  cycles are deformations of these, which may pass through operator insertions or their antipodes. As just discussed, these do not lead to monodromies. We conclude that the four point function is single-valued for all spatially winding cycles. 

Now, consider a four point function of operators of positions $\s_k, \bs_k$ for $k = 1, 2, 3, 4$. Exchanging $\bs_k \to -\bs_k$ for each operator sends $\sin\s_{ij}\sin\bs_{ij} \to -\sin\s_{ij}\sin\bs_{ij}$; hence, this transformation exchanges timelike and spacelike cycles and flips the sign of $\epsilon$.  As the resulting correlation function has been shown to be free of monodromies around spacelike cycles, the original correlation function must have no monodromies around timelike cycles. 
  
Contours with any winding numbers can be obtained by sewing copies of these generating cycles and therefore are also monodromy-free. 
  
 We conclude that the four point function of any CFT$_2$ has a globally defined extension from \m2\ to \et2.

\section{Beyond 4 points}
For n$>$4 points the correlator is a function of the ${n(n-1)(n-2)(n-3) \over 24}$ SO$(2,2)$-invariant  left and right cross-ratios $r_l,\bar r_k$. These take  the form of \eqref{rdef}, \eqref{brdef} for any subset of four points, and are themselves single-valued.  In adding a fifth operator  to a 4 point correlator, the argument of the preceding section implies that there is  no monodromy of the correlator as the 5th point is taken around any of the others. Similarly the known single-valuedness of the extension from \m2\ to  EC$^2$ together with the simple action of $\bs_k\to -\bs_k$ insure the absence of  monodromy around all cycles. One thereby iteratively deduce the n-pooint correlator is single-valued.

 \section{Higher dimensions}

In this section we consider  the extension to $D>2$ spacetime   dimensions. We will assume that  branch cuts in correlators  arise only when a pair of operators are null separated, which has been proven only for the special case $D=2$ \cite{Maldacena:2015iua}. Given this strong assumption we show that any such CFT$_D$ four point function has a continuation to the equal radii $S^{D-1}\times S^1$ Einstein torus ET$^D$.

\subsection{The embedding formalism} The correlators of a CFT$_D$ in M$^D$  are efficiently described in the embedding formalism.\footnote{For a recent reviews focusing on  Lorentzian CFT$_D$, see sections in \cite{Hartman:2015lfa,HarrisSimmonsDuffinTASI,Kundu:2025jsm}. The embedding formalism is also very efficient for spinning operators  \cite{Costa:2011mg}.} This begins in signature ($D,2$) flat space with coordinates $X^A$, 
$A=-1,0,\ldots,D$ and metric
\be ds_{4,2}^2=-(dX^{-1})^2+\eta_{\mu\nu}dX^\mu dX^\nu+(dX^D)^2,~~~\mu,\nu= 0, 1, \ldots D-1 \ee
on which $SO(D,2)$ acts linearly as the Lorentz group.  The projective light cone of the origin
\be \label{plc} X^2=0, ~~~~X^A \sim \lambda X^A, ~~~~\lambda >0, \ee
 then defines $D$-dimensional space on which SO$(D,2)$ acts as the conformal group. Any global section is conformal to ET$^D$. To recover Minkowski space, we take the  `Weyl frame'
 \be \label{sct} X^{-1}+X^D=1\ee
 known as the Poincare section. 
 The restriction $X^2=0$ implies
 \be X^A=({1+x^2 \over 2}, x^\mu,{1-x^2 \over 2}).\ee
 The induced metric on this section 
 \be ds^2=\eta_{\mu\nu}dx^\mu dx^\nu \ee
 is the flat metric on M$^D$.  A change of section replacing \eqref{sct} by
\be \label{ssct} X^{-1}+X^D=\Omega(x) \ee
leads to a Weyl transformation on the induced metric 
 \be\label{wf} ds^2=\Omega^2 (x)\eta_{\mu\nu}dx^\mu dx^\nu.\ee
 The special case of constant  $\Omega=\lambda_0 $  is a dilation.

 The constraints on CFT in M$^D$ imply that the correlators, when lifted to embedding space, must transform covariantly under both SO$(D,2)$ and Weyl transformations.  We begin with the two-point function between scalar operators at $x^\mu_1$ and $x^\mu_2$.  The only
 SO$(D,2)$ invariant function of the coordinates  is \be X_1\cdot X_2=-{(x_{12})^2 \over 2},\ee  where $x^\mu_{12}=x^\mu_{1}-x^\mu_{2}$.  An operator $\co^\D$ of dimension $\D$ is one which scales like $\lambda_0^{-\D}$ under dilations.  SO$(D,2)$ then  implies   that  the appropriately normalized two-point function, with $i\e$ prescription  for the time-ordered product, is \be\label{mcr}\langle \cO^\D(x_1)  \cO^\D(x_2) \rangle_{M_D} ={1 \over ((x_{12})^{2}+i\e)^\D},\ee
 with branch cut phases defined as in \eqref{bcp}. Under a change of Weyl frame \eqref{wf}
 \be\label{mcb}\langle \cO^\D(x_1)  \cO^\D(x_2) \rangle_{\Omega^2M_D} = \Omega^{-\D}(x_1)\Omega^{-\D}(x_2)\langle \cO^\D(x_1)  \cO^\D(x_2) \rangle_{M_D}.\ee
Similarly, scalar $n$-point functions depend only on the ${n(n-1)} \over 2$ invariants
 $X_k\cdot X_l$ and transform covariantly under \eqref{wf}. Spinning correlators involve polarization vectors \cite{Costa:2011mg}.  
 
 The projective light cone defined in \eqref{plc} is not fully covered by the Poincare section \eqref{sct}.  Full coverage may be obtained with  a second Poincare section  with 
 \be \label{ssct} X^{-1}+X^D=-1.\ee
 A single global section, covering both Minkowski diamonds,  is defined by 
 \be\label{ets} (X^{-1})^2+(X^{0})^2=1. \ee
 Restriction to the light cone $X^2=0$ then implies
% \be  (dX^{1})^2+(dX^{2})^2+ (dX^{3})^2 + \cdots +  (dX^{D})^2=0.\ee
\be (X^1)^2+(X^2)^2 + \cdots (X^D)^2 = 1.\ee
 The topology of this section is $S^{D-1}\times S^1$, and the null geodesics all close after a single circuit around the $S^1$. 
 
 For concreteness, consider the case $D=4$. Defining coordinates 
 \be\label{xet} X^A=\left (\sin t, \cos t, \sin \theta \sin \psi \cos \phi,\sin \theta \sin \psi \sin \phi, \sin \theta \cos \psi, \cos \theta \right) .\ee
one finds the induced metric of  ET$^4$
 \be ds_{ET_4}^2=-dt^2+d\theta^2+\sin^2\theta d\Omega_2^2,~~~t\sim t+2\pi.\ee
%with $d\Omega_{2}^2=\d\psi^2+\sin^2\psi d\phi^2$
The light cone of a point in this geometry  (say $t={\pi \over 2},~\theta=0$) initial expands outwards but then reconverges on the  other side of the sphere ($t={3\pi \over 2},~\theta=\pi$),  crosses itself and finally returns to its starting point. This light cone divides ET$^4$ into two causal diamonds each of which is conformal to flat M$^4$. To see this explicitly define new coordinates 
\be \label{sxa} T= {\cos t \over \Omega_0}~~~R={\sin \theta \over \Omega_0},~~~~\Omega_0=  \cos \theta -\sin t.\ee
One finds
\be \label{ctf}ds_{ET_4}^2=\Omega_0^{2}\big(-dT^2 +dR^2+R^2d\Omega_{2}^2\big),  \ee
which identifies $\Omega_0$ as the Weyl transformation relating ET$_4$ to two copies of M$^4$. 
The two Minkowski regions are distinguished  by the sign of  $\Omega_0$. 

\subsection{Conformal correlators on the Einstein torus}
Continuing the specialization to $D=4$, let's see how the M$^4$ correlator \eqref{mcr} is extended to all of ET$^4$. For a general section the   correlators \eqref{mcr} take the form  
\be\label{mzcr}\langle \cO^\D(x_1)  \cO^\D(x_2) \rangle ={1 \over (-2X_{1}\cdot X_2+i\e)^{\D}},\ee
Using \eqref{xet} one finds for the section\eqref{ets} that 
\be X_1\cdot X_2=-\cos t_{12}+\cos\theta_{12},\ee
where $\theta_{12}$ is the solid angle separating $X_1$ and $X_2$ on the $S^3$.  This implies that on ET$^4$
\be\label{mzcr}\langle \cO^\D(x_1)  \cO^\D(x_2) \rangle_{ET^4}={1 \over (2\cos t_{12}-2\cos\theta_{12}+i\e)^{\D}}.\ee
One may directly  check that (i) this is single-valued and (ii) performing a Weyl transformation $\Omega_0$ in either Minkowski diamond and the coordinate transformation \eqref{sxa} that this reduces to the original M$^4$ expression \eqref{mcr}. Therefore \eqref{mzcr} defines a continuation of any scalar CFT$_4$ Minkowski two-point function to ET$^4$. Similar constructions apply for general $D$.

We now argue that, given a reasonably-motivated assumption about where the four-point function has branch points, that this procedure defines a single-valued four-point function on the Einstein torus for general dimension $D3$. We focus on scalar operators, where the time-ordered M$^D$ four-point function can be written as 
\begin{equation}
    \langle \O_1(x_1)\cdots \O_4(x_4)\rangle = \mathcal{N}_4\ g(u,v),\ %u = \frac{x_{12}^2x_{34}^2}{x_{13}^2x_{24}^2},\ v = \frac{x_{14}^2x_{23}^2}{x_{13}^2x_{24}^2}
\end{equation}
where 
\be u=\frac{(x_{12}^2+i\eps)(x_{34}^2+i\eps)}{(x_{13}^2+i\eps)(x_{24}^2+i\eps)},  ~~~ v=\frac{(x_{14}^2+i\eps)(x_{23}^2+i\eps)}{(x_{13}^2+i\eps)(x_{24}^2+i\eps) },\ee  
$\mathcal{N}_4$ is a standard conformally covariant prefactor \cite{Rattazzi:2008pe}
\begin{equation}
    \mathcal{N}_4  = \left(\frac{x_{24}^2+i\eps}{x_{14}^2+i\eps}\right)^{\D_{12}/2}\left(\frac{x_{14}^2+i\eps}{x_{13}^2+i\eps}\right)^{\D_{34}/2}\frac{1}{(x_{12}^2+i\eps)^{(\D_1+\D_2)/2}(x_{34}^2+i\eps)^{(\D_3+\D_4)/2}}
\end{equation}
and $\D_{ij}=\D_i-\D_j$. Because spinning conformal blocks will have the same branch structure, our results will also hold for spinning operators \cite{Bissi:2019kkx}. 

Under our choice of analytic extension to ET$^D$, the conformally covariant prefactor becomes
\begin{equation}
\begin{split}
     \mathcal{N}_4 
     &= \left(\frac{-X_2\cdot X_4+i\eps}{-X_1\cdot X_4+i\eps}\right)^{\D_{12}/2}\left(\frac{-X_1\cdot X_5+i\eps}{-X_1\cdot X_3+i\eps}\right)^{\D_{34}/2}\\ &~~~~\times\frac{2^{(\D_1+\D_2+\D_3+\D_4)/2}}{(-X_1\cdot X_2+i\eps)^{(\D_1+\D_2)/2}(-X_3\cdot X_4+i\eps)^{(\D_3+\D_4)/2}}.
     \end{split}
\end{equation}
This can easily be seen to be a single-valued function of $X_i$ as $X_i\cdot X_j \pm i\epsilon$ has a fixed imaginary part and can never circle the origin \cite{Melton:2023hiq}. The conformal cross ratios extend to ET$^D$ by
\begin{equation}\label{eq:uvieps}
    u  = \frac{(-X_1\cdot X_2 + i\eps)(-X_3\cdot X_4+i\eps)}{(-X_1\cdot X_3+i\eps)(-X_2\cdot X_4+i\eps)},~~~~~ v  = \frac{(-X_1\cdot X_4 + i\eps)(-X_2\cdot X_3+i\eps)}{(-X_1\cdot X_3+i\eps)(-X_2\cdot X_4+i\eps)}.
\end{equation}
We now need to show that the function $g(u,v)$ is single-valued on the torus. To do so, we assume that the four-point function has branch cuts only when pairs of operators become null separated; i.e. where $u =  0$ ($x_{12}^2 = 0)$, $v = 0$ ($x_{14}^2 = 0$), or $u = v = \infty$ ($x_{13}^2 = 0$). While in 2D these are the only locations where a singularity can arise, in higher dimensions poles off of these locations are known to occur \cite{Maldacena:2015iua}. However, these do not lead to branch cuts of the conformal blocks for $D \neq  3$ \cite{Dolan:2003hv}.\footnote{In $D = 3$, the singularities in perturbative correlators described in \cite{Maldacena:2015iua} can be logarithmic, which is associated with the possibility of nonlocal anyonic CFTs which have monodromies even in the Minkowski patch. Additionally, in \cite{Hartman:2015lfa}, time delays leading to a more general branch structure when commutators of Lorentzian operators are considered. Given that our $i\epsilon$-prescription only samples time ordered correlation functions, we do not expect to be able to sample sheets  of the four-point function that exhibit these particular time delays. Nevertheless these examples indicate that the general validity of our assumption is not obvious for $D \neq 2$. Either a counterexample to or a proof of our assumption would be of great interest. }
\begin{figure}[h]
    \centering

\tikzset{every picture/.style={line width=0.75pt}} %set default line width to 0.75pt        

\begin{tikzpicture}[x=0.75pt,y=0.75pt,yscale=-1,xscale=1]
%uncomment if require: \path (0,300); %set diagram left start at 0, and has height of 300

%Shape: Axis 2D [id:dp7968003704930593] 
\draw  (57.86,169.39) -- (291.43,169.39)(181.05,66.1) -- (181.05,265.49) (284.43,164.39) -- (291.43,169.39) -- (284.43,174.39) (176.05,73.1) -- (181.05,66.1) -- (186.05,73.1)  ;
%Curve Lines [id:da6355319016988299] 
\draw    (182.05,176.39) .. controls (76.05,176.39) and (97.05,260.39) .. (180.05,260.39) ;
\draw [shift={(110.57,220.91)}, rotate = 264.2] [fill={rgb, 255:red, 0; green, 0; blue, 0 }  ][line width=0.08]  [draw opacity=0] (10.72,-5.15) -- (0,0) -- (10.72,5.15) -- (7.12,0) -- cycle    ;
%Curve Lines [id:da06510123206500473] 
\draw    (180.05,260.39) .. controls (289.05,265.39) and (287.05,160.39) .. (180,164.32) ;
\draw [shift={(260.59,207.49)}, rotate = 85.73] [fill={rgb, 255:red, 0; green, 0; blue, 0 }  ][line width=0.08]  [draw opacity=0] (10.72,-5.15) -- (0,0) -- (10.72,5.15) -- (7.12,0) -- cycle    ;
%Curve Lines [id:da5371617625547355] 
\draw    (180,164.32) .. controls (65.05,159.39) and (92.05,80.39) .. (183,82.2) ;
\draw [shift={(105.24,117.12)}, rotate = 101.95] [fill={rgb, 255:red, 0; green, 0; blue, 0 }  ][line width=0.08]  [draw opacity=0] (10.72,-5.15) -- (0,0) -- (10.72,5.15) -- (7.12,0) -- cycle    ;
%Curve Lines [id:da5498897976998836] 
\draw [color={rgb, 255:red, 0; green, 0; blue, 0 }  ,draw opacity=1 ]   (183,82.2) .. controls (297.05,84.39) and (262.05,178.39) .. (182.05,176.39) ;
\draw [shift={(254.82,132.98)}, rotate = 279.65] [fill={rgb, 255:red, 0; green, 0; blue, 0 }  ,fill opacity=1 ][line width=0.08]  [draw opacity=0] (10.72,-5.15) -- (0,0) -- (10.72,5.15) -- (7.12,0) -- cycle    ;
%Curve Lines [id:da23294848214747943] 
\draw    (391.05,164.39) .. controls (534.05,137.13) and (543.5,249) .. (448.5,262) ;
\draw [shift={(504.25,188.21)}, rotate = 232.94] [fill={rgb, 255:red, 0; green, 0; blue, 0 }  ][line width=0.08]  [draw opacity=0] (10.72,-5.15) -- (0,0) -- (10.72,5.15) -- (7.12,0) -- cycle    ;
%Curve Lines [id:da9463756994813045] 
\draw    (448.5,262) .. controls (385.5,267) and (306.5,179) .. (452.5,179) ;
\draw [shift={(369.95,209.04)}, rotate = 90.24] [fill={rgb, 255:red, 0; green, 0; blue, 0 }  ][line width=0.08]  [draw opacity=0] (10.72,-5.15) -- (0,0) -- (10.72,5.15) -- (7.12,0) -- cycle    ;
%Curve Lines [id:da13393594845587709] 
\draw    (452.5,179) .. controls (552.05,146.13) and (502.05,85.39) .. (451.05,85.39) ;
\draw [shift={(509.3,123.98)}, rotate = 85.36] [fill={rgb, 255:red, 0; green, 0; blue, 0 }  ][line width=0.08]  [draw opacity=0] (10.72,-5.15) -- (0,0) -- (10.72,5.15) -- (7.12,0) -- cycle    ;
%Curve Lines [id:da8469476150076085] 
\draw    (451.05,85.39) .. controls (364.05,85.39) and (351.05,179.39) .. (391.05,164.39) ;
\draw [shift={(381.78,116.24)}, rotate = 309.74] [fill={rgb, 255:red, 0; green, 0; blue, 0 }  ][line width=0.08]  [draw opacity=0] (10.72,-5.15) -- (0,0) -- (10.72,5.15) -- (7.12,0) -- cycle    ;
%Shape: Axis 2D [id:dp35903456800333344] 
\draw  (325.86,168.39) -- (559.43,168.39)(449.05,65.1) -- (449.05,264.49) (552.43,163.39) -- (559.43,168.39) -- (552.43,173.39) (444.05,72.1) -- (449.05,65.1) -- (454.05,72.1)  ;
%Straight Lines [id:da011914740668940715] 
\draw    (180,164.32) ;
\draw [rotate = 0] [color={rgb, 255:red, 0; green, 0; blue, 0 }  ][fill={rgb, 255:red, 0; green, 0; blue, 0 }  ][line width=0.75]      (0, 0) circle [x radius= 3.35, y radius= 3.35]   ;
%Shape: Circle [id:dp93989567214195] 
\draw  [fill={rgb, 255:red, 126; green, 211; blue, 33 }  ,fill opacity=1 ] (177.99,164.28) .. controls (177.99,166.07) and (179.44,167.51) .. (181.23,167.51) .. controls (183.02,167.51) and (184.47,166.07) .. (184.47,164.28) .. controls (184.47,162.49) and (183.02,161.04) .. (181.23,161.04) .. controls (179.44,161.04) and (177.99,162.49) .. (177.99,164.28) -- cycle ;
%Shape: Circle [id:dp8585712304272317] 
\draw  [fill={rgb, 255:red, 208; green, 2; blue, 27 }  ,fill opacity=1 ] (177.99,82.28) .. controls (177.99,84.07) and (179.44,85.51) .. (181.23,85.51) .. controls (183.02,85.51) and (184.47,84.07) .. (184.47,82.28) .. controls (184.47,80.49) and (183.02,79.04) .. (181.23,79.04) .. controls (179.44,79.04) and (177.99,80.49) .. (177.99,82.28) -- cycle ;
%Shape: Circle [id:dp4405348666605585] 
\draw  [fill={rgb, 255:red, 144; green, 19; blue, 254 }  ,fill opacity=1 ] (177.99,177.28) .. controls (177.99,179.07) and (179.44,180.51) .. (181.23,180.51) .. controls (183.02,180.51) and (184.47,179.07) .. (184.47,177.28) .. controls (184.47,175.49) and (183.02,174.04) .. (181.23,174.04) .. controls (179.44,174.04) and (177.99,175.49) .. (177.99,177.28) -- cycle ;
%Shape: Circle [id:dp060730228850621626] 
\draw  [fill={rgb, 255:red, 0; green, 0; blue, 0 }  ,fill opacity=1 ] (177.99,260.28) .. controls (177.99,262.07) and (179.44,263.51) .. (181.23,263.51) .. controls (183.02,263.51) and (184.47,262.07) .. (184.47,260.28) .. controls (184.47,258.49) and (183.02,257.04) .. (181.23,257.04) .. controls (179.44,257.04) and (177.99,258.49) .. (177.99,260.28) -- cycle ;
%Shape: Circle [id:dp6524068335035528] 
\draw  [fill={rgb, 255:red, 126; green, 211; blue, 33 }  ,fill opacity=1 ] (445.99,160.28) .. controls (445.99,162.07) and (447.44,163.51) .. (449.23,163.51) .. controls (451.02,163.51) and (452.47,162.07) .. (452.47,160.28) .. controls (452.47,158.49) and (451.02,157.04) .. (449.23,157.04) .. controls (447.44,157.04) and (445.99,158.49) .. (445.99,160.28) -- cycle ;
%Shape: Circle [id:dp7055341754004267] 
\draw  [fill={rgb, 255:red, 0; green, 0; blue, 0 }  ,fill opacity=1 ] (445.99,262.28) .. controls (445.99,264.07) and (447.44,265.51) .. (449.23,265.51) .. controls (451.02,265.51) and (452.47,264.07) .. (452.47,262.28) .. controls (452.47,260.49) and (451.02,259.04) .. (449.23,259.04) .. controls (447.44,259.04) and (445.99,260.49) .. (445.99,262.28) -- cycle ;
%Shape: Circle [id:dp2036217824046731] 
\draw  [fill={rgb, 255:red, 144; green, 19; blue, 254 }  ,fill opacity=1 ] (444.99,178.28) .. controls (444.99,180.07) and (446.44,181.51) .. (448.23,181.51) .. controls (450.02,181.51) and (451.47,180.07) .. (451.47,178.28) .. controls (451.47,176.49) and (450.02,175.04) .. (448.23,175.04) .. controls (446.44,175.04) and (444.99,176.49) .. (444.99,178.28) -- cycle ;
%Shape: Circle [id:dp47216764297036407] 
\draw  [fill={rgb, 255:red, 208; green, 2; blue, 27 }  ,fill opacity=1 ] (445.99,85.28) .. controls (445.99,87.07) and (447.44,88.51) .. (449.23,88.51) .. controls (451.02,88.51) and (452.47,87.07) .. (452.47,85.28) .. controls (452.47,83.49) and (451.02,82.04) .. (449.23,82.04) .. controls (447.44,82.04) and (445.99,83.49) .. (445.99,85.28) -- cycle ;
%Straight Lines [id:da19778382758062352] 
\draw    (60,170) .. controls (61.66,168.33) and (63.33,168.32) .. (65,169.97) .. controls (66.67,171.63) and (68.34,171.62) .. (70,169.95) .. controls (71.66,168.28) and (73.33,168.27) .. (75,169.92) .. controls (76.67,171.58) and (78.34,171.57) .. (80,169.9) .. controls (81.66,168.23) and (83.33,168.22) .. (85,169.87) .. controls (86.67,171.53) and (88.34,171.52) .. (90,169.85) .. controls (91.66,168.18) and (93.33,168.17) .. (95,169.82) .. controls (96.67,171.48) and (98.34,171.47) .. (100,169.8) .. controls (101.66,168.13) and (103.33,168.12) .. (105,169.77) .. controls (106.67,171.43) and (108.34,171.42) .. (110,169.75) .. controls (111.66,168.08) and (113.33,168.07) .. (115,169.72) .. controls (116.67,171.38) and (118.34,171.37) .. (120,169.7) .. controls (121.66,168.03) and (123.33,168.02) .. (125,169.67) .. controls (126.67,171.33) and (128.34,171.32) .. (130,169.65) .. controls (131.66,167.98) and (133.33,167.97) .. (135,169.62) .. controls (136.67,171.28) and (138.34,171.27) .. (140,169.6) .. controls (141.66,167.93) and (143.33,167.92) .. (145,169.57) .. controls (146.67,171.23) and (148.34,171.22) .. (150,169.55) .. controls (151.66,167.88) and (153.33,167.87) .. (155,169.52) .. controls (156.67,171.18) and (158.34,171.17) .. (160,169.5) .. controls (161.66,167.83) and (163.33,167.82) .. (165,169.47) .. controls (166.67,171.13) and (168.34,171.12) .. (170,169.45) .. controls (171.66,167.78) and (173.33,167.77) .. (175,169.42) .. controls (176.67,171.08) and (178.34,171.07) .. (180,169.4) -- (181.05,169.39) -- (181.05,169.39) ;
%Straight Lines [id:da33850911264567285] 
\draw    (328,169) .. controls (329.66,167.33) and (331.33,167.32) .. (333,168.97) .. controls (334.67,170.63) and (336.34,170.62) .. (338,168.95) .. controls (339.66,167.28) and (341.33,167.27) .. (343,168.92) .. controls (344.67,170.58) and (346.34,170.57) .. (348,168.9) .. controls (349.66,167.23) and (351.33,167.22) .. (353,168.87) .. controls (354.67,170.53) and (356.34,170.52) .. (358,168.85) .. controls (359.66,167.18) and (361.33,167.17) .. (363,168.82) .. controls (364.67,170.48) and (366.34,170.47) .. (368,168.8) .. controls (369.66,167.13) and (371.33,167.12) .. (373,168.77) .. controls (374.67,170.43) and (376.34,170.42) .. (378,168.75) .. controls (379.66,167.08) and (381.33,167.07) .. (383,168.72) .. controls (384.67,170.38) and (386.34,170.37) .. (388,168.7) .. controls (389.66,167.03) and (391.33,167.02) .. (393,168.67) .. controls (394.67,170.33) and (396.34,170.32) .. (398,168.65) .. controls (399.66,166.98) and (401.33,166.97) .. (403,168.62) .. controls (404.67,170.28) and (406.34,170.27) .. (408,168.6) .. controls (409.66,166.93) and (411.33,166.92) .. (413,168.57) .. controls (414.67,170.23) and (416.34,170.22) .. (418,168.55) .. controls (419.66,166.88) and (421.33,166.87) .. (423,168.52) .. controls (424.67,170.18) and (426.34,170.17) .. (428,168.5) .. controls (429.66,166.83) and (431.33,166.82) .. (433,168.47) .. controls (434.67,170.13) and (436.34,170.12) .. (438,168.45) .. controls (439.66,166.78) and (441.33,166.77) .. (443,168.42) .. controls (444.67,170.08) and (446.34,170.07) .. (448,168.4) -- (449.05,168.39) -- (449.05,168.39) ;

% Text Node
\draw (101,41.4) node [anchor=north west][inner sep=0.75pt]    {$\textcolor[rgb]{0.49,0.83,0.13}{(}\textcolor[rgb]{0.49,0.83,0.13}{2,+}\textcolor[rgb]{0.49,0.83,0.13}{)}\textcolor[rgb]{0.82,0.01,0.11}{(}\textcolor[rgb]{0.82,0.01,0.11}{3,+}\textcolor[rgb]{0.82,0.01,0.11}{)}\textcolor[rgb]{0.56,0.07,1}{(}\textcolor[rgb]{0.56,0.07,1}{2,-}\textcolor[rgb]{0.56,0.07,1}{)}( 3,-)$};
% Text Node
\draw (370,41.4) node [anchor=north west][inner sep=0.75pt]    {$\textcolor[rgb]{0.49,0.83,0.13}{(}\textcolor[rgb]{0.49,0.83,0.13}{2,+}\textcolor[rgb]{0.49,0.83,0.13}{)}( 3,-)\textcolor[rgb]{0.56,0.07,1}{(}\textcolor[rgb]{0.56,0.07,1}{2,-}\textcolor[rgb]{0.56,0.07,1}{)}\textcolor[rgb]{0.82,0.01,0.11}{(}\textcolor[rgb]{0.82,0.01,0.11}{3,+}\textcolor[rgb]{0.82,0.01,0.11}{)}$};

\end{tikzpicture}
    \caption{Trajectories in the complex $u_R$ plane. No matter the order in which the operator 1 moves past the lightcones of operators 2 and 3, $u$ never circles the origin with this $i\epsilon$ prescription. The curve always passes the real axis at $u_R = 1$. The colored dots show the location of $u_R$ each time $\O_1$ moves across $\O_2$'s or $\O_3$'s lightcone.}
    \label{fig:trajplot}
\end{figure}
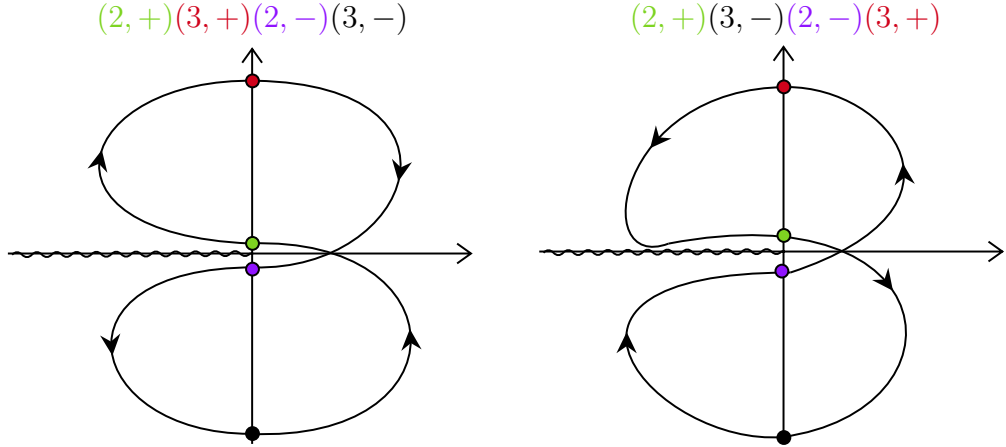
Consider fixing (generic) $X_{2,3,4}$ and taking $X_1$ around a spacelike or timelike cycle. For simplicity, we focus on the reduced variable 
\begin{equation}
    u_R = \frac{-X_1\cdot X_2 + i\eps}{-X_1\cdot X_3+i\eps},
\end{equation}
as the additional factor is simply a fixed real number for fixed $X_{2,3,4}$. As $X_1$ is moved around the timelike cycle, $u_R$ will trace out a trajectory in the complex plane.  $g(u,v)$ can have branch points if this trajectory encircles the origin  $u_R = 0$. To see that this is impossible we note that the imaginary part of $u_R$ vanishes only when $X_1 \cdot X_2=X_1 \cdot X_3$, whcih implies $u_R=1$. Therefore no trajectory of $u_R$ can ever encircle the origin; since $u$ is simply a rescaling of $u_R$ for fixed $X_{2,3,4}$, this implies that $u$ can never encircle $u = 0$ or $u = \infty$ when $X_1$ is moved around any cycle. An identical argument shows that $v$ can never circle $v = 0$ or $v = \infty$ as well, implying that the four-point function is single valued provded that $g$ only has branch cuts at $u = 0$, $v = 0$, or $u = v = \infty$.

It is instructive to see in detail how all the $u$ avoids circling 0 or $\infty$ when light cones are passed around a closed cycle. As $X^1$ is taken around a full timelike cycle
it will pass through operator 2's lightcone and operator 3's lightcone exactly twice. Let $(i,\pm)$ denote a crossing where $X_1$ crosses $X_i \cdot X_1 = 0$  with $X_1\cdot X_i$ increasing (decreasing). We can then represent any  cycle by listing the order in which $X_1$ crosses these lightcones. We now show that for any possible order of crossings that $u_R$ never circles the branch points at $u_R = 0$ or $u_R = \infty$.

First, consider a path labelled by $(2,\pm)(2,\mp)(3,\pm')(3,\mp')$. In this case, $u_R$ will pass by 0 with fixed sign of $X_1\cdot X_3$ and $u_R$ will pass by $\infty$ with fixed sign of $X_1 \cdot X_2$, so $u_R$ will move by 0 or $\infty$ and then return on the same side of the branch point. Hence the correlation function will have vanishing monodromy arouund this path.

Up to cyclic permutations, the remaining paths we need to check are $(2,+)(3,+)(2,-)(3,-)$ and $(2,+)(3,-)(2,-)(3,+)$. In the first case, $u_R$ starts out on the positive real axis, passes underneath $u = 0$ to the left, swings around to large negative $\mathrm{Im}\ u_R$, passes above $u_R = 0$ to the left, and then passes large positive $\mathrm{Im}\ u_R$ to return to its original point. Hence, it will never fully encircle the origin. In the second case, $u_R$ starts out on the negative real axis, passes above $u_R = 0$ to the right, moves to large negative $\mathrm{Im}\ u_R$, passes below $u_R = 0$ to the right, and then moves through large positive $\mathrm{Im}\ u_R$ to return to its starting point. These trajectories are depicted in Figure \ref{fig:trajplot}.

$v$ always takes an analogous path depending on how $X_1$ crosses the lightcone of operators 3 and 4. As such, provided that $g$ only has branch cuts at $u = 0, \infty$ and $v = 0, \infty$, where pairs of particles become null separated, the four-point function with $i\eps$ prescription given by Equation \eqref{eq:uvieps} has no monodromies around spacelike or timelike cycles. 

Higher point correlation functions can be written as a conformally covariant prefactor times a function of conformal cross ratios constructed from any four points. Provided that higher point correlation functions develop branch cuts only where pairs of operators become null separated, a similar  analysis will imply that the higher point correlation functions are also single-valued on ET$^D$.
 \section{Discussion}
  In addition to providing a new natural mathematical setting for studies of  CFT$_D$, the existence of   correlators on ET$^D$ is of potential interest for several reasons:
 
% \noindent(i) It provides a mathematically natural new  laboratory for the study of Lorentzian CFT$^D$ which, unlike the Einstein cylinder and Minkowski space, admits  a canonical  action of the conformal group $SO(D,2)$.
\begin{itemize}
\item[(i)] The result applies to the holographic CFT$_2$ duals appearing in string theory. It thereby allows us to define string theory on AdS$_3/\mathbb{Z}$ with closed timelike curves  via boundary correlators. These will have a $T$-dual representation along the timelike circles. Timelike  $T$-dual string theories in $M^{10}$ were studied   by Hull \cite{Hull:1998vg}. These  theories involve some unusual  signs and factors of $i$, but must be well-defined in the AdS$_3/\mathbb{Z}$ context. Moreover, they  contain spacelike $D$-branes and may  provide an interesting laboratory for timelike holography.

\item[(ii)] This work was in part inspired by investigations in celestial holography, in particular of leaf correlators. These  are CFT$_2$ correlators living on the ET$^2$ boundary of the 
AdS$_3/\mathbb{Z}$ leaves of a hyperbolic foliation of flat $(2,2)$ Klein space \cite{Atanasov:2021oyu,Melton:2023bjw,Melton:2024akx}. The leaf correlators are smooth objects defined by the (AdS$_3/\mathbb{Z}$)/CFT$_2$ dictionary \cite{Melton:2025ecj} and provide  building blocks of the full celestial correlators. The linear combinations which reassemble the celestial correlators nontrivially exhibit the distributional features required by spacetime translation invariance \cite{Melton:2023bjw,Melton:2024akx}.  Self-consistency of this construction requires the existence of a leaf  CFT$_2$ on the \et2 boundary of AdS$_3/\mathbb{Z}$ . It was this observation that led us to suspect that  CFT$_2$ correlators might generically  be defined on ET$^2$.

%In \cite{Melton:2023hiq}, the general kinematic form of a CFT$_2$ correlators on ET$^2$  was found.  In \cite{Melton:2023bjw}, it was shown that the relevant linear combinations of these smooth correlation functions, which at three points resemble Equation \eqref{cm5}, generate the distributional celestial amplitudes. These leaf amplitudes can be realized by an explicit  CFT$_2$ in Euclidean signature \cite{Melton:2024akx}. This work suggests that the leaf CFT can be intrinsically defined on the celestial torus. 

The companion paper \cite{Melton:2025ecj} defined a consistent  geometric quantization of free QFT on AdS$_3/\mathbb{Z}$. %It was shown that on the unitary principal series - commonly employed in celestial holography - there is a positive norm on the Hilbert space.  
This  work is the bulk counterpart of the boundary ET$_2$ analysis presented here. 
However, enabled by the powerful methods of CFT, the current paper  goes a step further with the inclusion of interactions. 

\item[(iii)] Our work constructs a large family of non-trivial self-consistent interacting quantum  systems on  spacetimes with closed timelike curves. There is a considerable literature on this subject (see e.g. \cite{Deutsch:1991nm,Hartle:1993sg,Bennett:2009rt,Lloyd:2010nt,Luminet:2021qae,Bishop:2024cqa}) for which this work may provide useful examples. 
\end{itemize}

\section*{Acknowledgements}
 
This work was supported by DOE grant de-sc/0007870, the Simons Collaboration on Celestial Holography and the Harvard Society of Fellows. We are grateful to Matthew Dodelson, Tom Hartman, Simon Hueveline, Lionel Mason, Shiraz Minwalla, Romain Ruzziconi, Atul Sharma, David Simmons-Duffin, Tianli Wang and Xi Yin for useful conversations.

\bibliographystyle{JHEP}
\bibliography{refs}

\end{document}